\documentstyle[11pt,pasp,twoside]{article}
\def\edcomment#1{\iffalse\marginpar{\raggedright\sl#1\/}\else\relax\fi}
\reversemarginpar
\begin{document}
\title{Thermal Limit Spectroscopy as a Goal for X-ray Astronomy}
\author{Martin Elvis} 
\affil{Harvard-Smithsonian Center for Astrophysics, 60 Garden
St., Cambridge MA 02138 USA}

\begin{abstract}

The $R\sim$300--1000 grating spectra from XMM-Newton and {\em Chandra}
are a radical advance, allowing spectroscopic physics techniques
to be applied to X-ray astronomy, revolutionizing a wide range of
research. Ten years on these spectra will be routine, and higher
resolution will be needed.  I propose ``Thermal Limit
Spectroscopy'' as the next natural goal for X-ray
spectroscopy. This will open up new physics: plasma physics,
velocity widths, Doppler shifts, line profiles, and absorption
lines in photoionized plasmas.  A resolution of $R$=3000--10,000
is required, and the technology is within reach.

\end{abstract}

\section{Introduction: the Near Future in X-ray Spectroscopy}

The grating spectra from {\em Chandra} and XMM-Newton are a startling
change from the non-dispersive spectra of the earlier era.  At a
resolution of $\sim$300-- $\sim$1000 they clearly separate out
many the emission lines in soft X-ray spectra for the first time,
and allow the application of the physics based techniques of
spectroscopy to X-ray astronomy, revolutionizing a vast range of
research areas.  {\em Chandra} and XMM-Newton spectra already show that
ten years from now we will be in great need of higher resolution
spectra.

The next generation of X-ray astronomy missions, ASTRO-E~II,
Con-X and XEUS, are all primarily spectroscopic, and share the
goal of high signal-to-noise, trading better resolution for
larger collecting area.  All three missions primarily use
calorimeters, which should reach $\sim$1-2~eV resolution by
launch (Stahle, these proceedings), i.e. $R\sim$500-1000 at
1~keV, similar to {\em Chandra} or XMM-Newton.  The early {\em Chandra} HETG
spectra in particular, show a great abundance of atomic features
below about 1.5~keV, as expected from atomic physics and cosmic
abundances, while only the Fe-K line is prominent above
$\sim$3~keV.  So 1~keV (12.54\AA) makes a good reference point
for resolution.

How much more resolution do we need?  I propose ``Thermal Limit
Spectroscopy'' as the next natural goal of X-ray spectroscopy
(Elvis \& Fabbiano 1997).  The thermal width separates the
temperature of the ions ($T_{ion}$, e.g. O$^{5+}$) from the
electron temperature ($T_e$); velocity widths, Doppler shifts,
line profiles, and absorption lines in photoionized plasmas all
then become accessible.
A kT=1~keV plasma has thermal velocities of
$\sim$100~km~s$^{-1}$, requiring a resolution of
$R$=3000--10,000. The technology for reaching this regime is
within reach.

\section{A Major Step: Thermal Limit Spectroscopy}

The goal of $R\sim$1000 was driven by a happy combination of
scientific goals (resolving the Helium-like triplets) and
technological capability (good mirrors and gratings).  The next
natural goal for X-ray spectroscopy is thermal limit spectroscopy
at $R\sim$10,000, because the next level of physics, as optical
and ultraviolet astronomy shows, comes from line profiles.  This
goal is not technologically absurd. The XMM-Newton RGS gratings
would give $R\sim$6500 if placed behind the $\frac{1}{2}$arcsec
{\em Chandra} mirrors
\footnote{Albeit with a new mounting scheme. The present grating
mounts were designed to work with the XMM-Newton mirrors, and so
are not controlled to a level that would let their inherent
resolution show. According to the RGS designers this is not a
major engineering challenge.}
.

The typical temperature of an X-ray hot plasma emitting at 1~keV
is $T\sim$10$^7$K (i.e. $kT$=1~keV, naturally).  For a thermal
plasma in coronal equilibrium at $T\sim$10$^7$K the typical
mid-atomic number (Z) elements such as Oxygen have thermal
velocities of $v_{th}\sim$125~km~s$^{-1}$. (A proton has
$v_{th}\sim$500~km~s$^{-1}$, while an iron atom
$v_{th}\sim$75~km~s$^{-1}$.) A resolution of 3000 thus gives
1~pixel/line~width, and a resolution of 10,000 gives a good
oversampling of 3~pixels/line~width
\footnote{The natural, uncertainty principle, line width
$\gamma=\hbar A$ (where $A$ is the Einstein A coefficient) , is
almost always 0.001-0.1~eV, at most comparable with the thermal
width in X-ray hot plasmas, since $10^{12}<A<10^{14}$~s$^{-1}$.}
. Reaching this resolution brings qualitative improvements in the
science we can extract from our data, improvements that are not
at all restricted to thermal plasmas. I outline some of these
below.

\smallskip
\noindent\underline{\em More, different lines.}
A factor of 2 more lines will be separated out of blends at
$R$=10,000 vs. $R$=300 (Smith \& Brickhouse 2000).  More
importantly a different type of line becomes resolved at
$R$=10,000: dielectronic recombination lines. These are always
close to the resonance lines and are always blended at $R$=1000,
but can be resolved at $R$=10,000. The dielectronic recombination
lines depend only on $T_e$ and ionization, so providing clear new
diagnostics.

\smallskip
\noindent\underline{\em Thermal Plasma Physics.}
Normally we assume that cosmic thermal plasmas are equilibrated,
i.e. that $T_e$ = $T_{ion}$. Often this is not a good assumption,
e.g. any rapidly expanding plasma (solar wind, SNR, galaxy
outflows).  Testing ion temperatures with line widths
\footnote{
$\propto Z^{-\frac{1}{2}}$, c.f. constant width for turbulence
}
vs. $T_e$ is a basic measurement enabled by $R\sim$10,000.
For decades the heating mechanism of the solar corona has been a
puzzle. SOHO UVCS has now shown that $T_{ion} > T_e$, ruling out
Ohmic heating (for which $T_{ion} < T_e$), and suggesting initial
Alfven heating of the ions (Cranmer 2000).  Are all stellar
coronae heated this way?

\smallskip
\noindent\underline{\em Photoionized Plasmas.}
Thermal plasmas cooler than $\sim$10$^5$K do not generally
produce X-ray lines. Cool {\em photoionized} plasmas though can
show X-ray transitions at all temperatures. Cool material
commonly surrounds X-ray sources: X-ray binaries, HII regions and
AGN are all rich in cool gas. Measuring these lines will tell us
about the flows of this gas, and hence dynamics and mass loss
rates, and can delimit the often unseen ionizing
continuum. Radiatively driven winds, (e.g. P~Cygni) are unstable
and typically break up, producing many narrow
absorption features. While narrow emission lines are easily
detected, an absorption line from 10$^4$K gas (thermal width
$\sim$10~km~s$^{-1}$) needs $R\sim$15,000.

\smallskip
\noindent\underline{\em Doppler Motions.}
The Doppler effect is one of the most employed methods for
understanding astronomical objects. In highly ionized matter
around X-ray sources optical or ultraviolet lines are weak, while
X-ray transitions dominate. Systems with velocities
$\sim$1000~km~s$^{-1}$ are open to {\em Chandra} and
XMM-Newton. Most Doppler velocities are over an order of
magnitude smaller: e.g. G-star winds, where {\em Doppler
tomography} via rotation can map coronal structures, have $\Delta
v \sim$10~km~s$^{-1}$; X-ray binaries have orbital velocities $v
\sim$300~km~s$^{-1}$; the hot ISM can map Galactic structure over
a range $v \sim$0--300~km~s$^{-1}$; AGN absorbers have $\Delta v
\sim$30~km~s$^{-1}$; and cluster mergers take place at the sound
speed $v \sim$300~km~s$^{-1}$.  To measure line centroids to
$\sim$20~km~s$^{-1}$ needs a FWHM$<$ 200~km~s$^{-1}$, i.e. $R
>$3000.

\smallskip
\noindent\underline{\em Photoabsorption Edge Resonances \& Dust.}
The dust-to-gas ratio is key to many astrophysical situations
(e.g. planet formation), but is surprisingly hard to
measure. X-ray edge structures (e.g. McLaughlin \& Kirby 1998)
depend on the molecular state of the atoms. e.g.  Is oxygen is
atomic, O$_2$, CO$_2$, or in a silicate dust particle such as
MgSiO$_3$.  Ferric ISM dust grains have been found this way with
{\em Chandra} (Paerels et al., 2001). Many edge structures
require $R>$1000.

\smallskip
\noindent\underline{\em The Killer App: The Missing Baryons.}
While compelling science, none of the above is readily condensed
into a sound bite that can capture public attention.  One
application could do: the study of the `missing baryons'.  At low
z most `normal' matter should be hidden in a warm-hot
(10$^6$K--10$^7$K) intergalactic medium studiable only in soft
X-rays (Fiore, these proceedings). {\em Chandra} will probably
give the first detection of this medium.  Detailed study though -
are they thermal or photoioized, engaged in large scale motions,
polluted by galaxy superwinds - requires resolving line widths
and looking for small Doppler shifts, for many ions. This
requires thermal limit X-ray spectroscopy.


\vspace{-2mm}
\section{Achieving Thermal Limit Spectroscopy}

Even if thermal limit spectroscopy is an obvious goal for X-ray
astronomy, it will not be accepted as the right next step if the
technology is too far from realization. Luckily, both mirrors and
spectrometers are surprisingly close to the needed performance.

\smallskip
\noindent{\bf Mirror Size.}
The power of any telescope depends on `$f$At$\epsilon$'. The
product of flux, area, exposure time and efficiency, when divided
by the mean photon energy, determines the number of photons
collected.
At high resolution emission lines need only slightly greater
mirror area, since they will simply show up more clearly against
a zero continuum value until resolved.
Narrow absorption lines need a minimum of 10 continuum counts per
bin to be detected significantly. At $R$=5000 this requires
1~sq.meters effective area for a source at
2$\times$10$^{-12}$erg~cm$^{-2}$~s$^{-1}$ (i.e. $\sim$3000
targets in the sky) in 10$^4$s. This is comparable to Con-X and
smaller than XEUS-I.

\smallskip
\noindent{\bf Spectrometers.}

\noindent\underline{\em Calorimeters} combine high efficiency
with a broad band and `integral field' spectroscopy (i.e. a
spectrum at every image pixel). However the thermal noise
requires that such a calorimeter operate at $\sim$5~mK,
vs. $\sim$50~mK for today's instruments, but other noise sources
may prevent its realization.

\noindent\underline{\em Bragg Crystals} naturally reach
$R\sim$1000 and have an intrinsically broad field of
view. However they have narrow ($\sim$1\%) band passes and so a
low efficiency if many lines are needed. Reaching $R\sim$10,000
is not easy.

\noindent\underline{\em Transmission Gratings} are lightweight
and cover a large bandwidth. They have $\epsilon$=0.3,
vs. $\epsilon\sim$1 for calorimeters, but this is small factor
compared with the increase in mirror area needed in any
case. Transmission grating periods of 80\AA, 5 times smaller than
on the {\em Chandra} MEG at 1~keV, are now feasible (Savas et al
1996), resulting in $R\sim$5000 behind a {\em Chandra}-like
mirror.  Extended sources are non-trivial however.  `Long-slit'
designs, as in optical astronomy, could overcome this problem, at
the cost of doubling the length of the optics system.  This may
no longer be a significant constraint now that station keeping
(e.g. XEUS) is needed.  Transmission gratings require a high
resolution mirror, improving $R$ as the mirror improves. Such
mirrors are needed in any case for imaging applications
(Fabbiano, these proceedings).

\noindent\underline{\em Reflection Gratings} can already provide
resolutions of the right order, given a $\sim$1~arcsecond
mirror. Again extended sources would require `long slit' optics.
`Out of plane' geometries (Cash 1991) offer higher spectral
resolution, and higher efficiency.  With 5$^{\prime\prime}$ HEW
mirrors Con-X could achieve $R$=5000 (Cash 2001).

\vspace{-4mm}
\section{Conclusion}

Cosmic X-ray spectroscopy is a new field. At last the techniques
of atomic physics are being applied to astronomical X-ray
sources. With $R\sim$1000 there are wonderful riches, but a
decade from now will that be enough? The promise of $R\sim$10,000
resolution is technologically not too distant.  Thermal limit
X-ray spectroscopy is the right choice of goal for a new
generation of X-ray missions.

\smallskip
This work was supported in part by NASA grant NAG5-6078.

\vspace{-4mm}


\end{document}